\documentclass[article]{revtex4}
\usepackage{graphicx}

\begin{document}
\title{Phase Transition in the Aldous-Shields Model of Growing Trees}

\author{David S. Dean $^{1}$ and Satya N. Majumdar $^{2}$}
\address{
{\small $^1$ Laboratoire de Physique Th\'eorique,  UMR CNRS 5152, IRSAMC, Universit\'e
Paul Sabatier, 118 route de Narbonne, 31062 Toulouse Cedex 04, France}\\
{\small $^2$ Laboratoire de Physique Th\'eorique et Mod\`eles Statistiques, UMR CNRS
8626,Universit\'e Paris Sud, B\^at 100, 91045 Orsay Cedex, France}
}

\date{17 October 2005}

\begin{abstract}
We study analytically the late time statistics of the number of
particles in a growing tree model introduced by Aldous and Shields. In
this model, a cluster grows in continuous time on a binary Cayley
tree, starting from the root, by absorbing new particles at the empty
perimeter sites at a rate proportional to $c^{-l}$ where $c$ is a
positive parameter and $l$ is the distance of the perimeter site from
the root. For $c=1$, this model corresponds to random binary search
trees and for $c=2$ it corresponds to digital search trees in computer
science. By introducing a backward Fokker-Planck approach, we
calculate the mean and the variance of the number of particles at
large times and show that the variance undergoes a `phase transition' at
a critical value $c=\sqrt{2}$. While for $c>\sqrt{2}$ the variance is
proportional to the mean and the distribution is normal, for
$c<\sqrt{2}$ the variance is anomalously large and the distribution is
non-Gaussian due to the appearance of  extreme fluctuations.
The model is generalized to one where growth occurs on a tree with $m$
branches and, in this more general case, we show that the critical
point occurs at $c=\sqrt{m}$.
\bigskip



\end{abstract}
\maketitle

\section{Introduction}

Growing clusters are ubiquitous in nature and they exhibit fascinating
structures and patterns.  Examples range from natural fractals, such
as snowflakes and soots, to artificial structures such as networks,
for example the Internet and social networks.  
Various growth models have been studied
extensively by physicists over the last three
decades~\cite{Growth}. In these models growth starts from a single
seed site and proceeds via absorbing new particles into the cluster
according to certain specified rules. Different growth rules give rise
to different growth models, examples being the Eden model~\cite{Eden},
invasion percolation~\cite{IP}, diffusion limited
aggregation~\cite{DLA} and the growing network models~\cite{Networks}
which have recently received much attention. There are two reasons
why many of these growth models are often studied on a Cayley tree (or
on the Bethe lattice)~\cite{Edentree}. First, the tree structure of
the Bethe lattice mimics a Euclidean lattice in the limit of high
dimensions where the mean field theory often becomes exact. Secondly,
the absence of loops on the Cayley tree often allows one to obtain
exact analytical solutions which are very difficult to obtain on a
regular $d$-dimensional lattice. There is yet another compelling
motivation for studying these growth models on a Cayley tree and this
comes from computer science. `Storing and Search' of data is a very
important area of computer science~\cite{Knuth}. Incoming data to a
computer is usually stored on a Cayley tree by using various data
storage algorithms and the tree so grown is called a `search
tree'~\cite{Knuth}. Different algorithms lead to different search
trees and in some cases, as explained below, the rules of growth of a
search tree can be shown to be exactly equivalent to a `physical'
growth model on the tree. Thus the study of these physical growth
models on a Cayley tree provides important insights into data storage
in computer science.

As a first example of this equivalence between a physical growth model
and a search tree, we show here that the Eden model on a binary Cayley
tree is exactly equivalent to the random binary search tree
(RBST). Consider the Eden model on a binary Cayley tree where the
growth starts from the root~\cite{Edentree}. At the first step, a
particle gets absorbed at the root, thus forming a cluster of size
$1$. This cluster has now two empty neighbors which defines the
perimeter of the cluster. At the next step, a new particle will get
absorbed at any of these two perimeter sites chosen with equal
probability, thus forming a cluster of size $2$. The subsequent growth
occurs following the same rule, namely a new particle gets absorbed at
any of the perimeter sites chosen with equal probability. In Fig. (1),
we show a cluster after $4$ steps where the black sites denote the
cluster and the shaded sites denote the current perimeter sites that
are available for subsequent growth.
\begin{figure}
\includegraphics[width=4in]{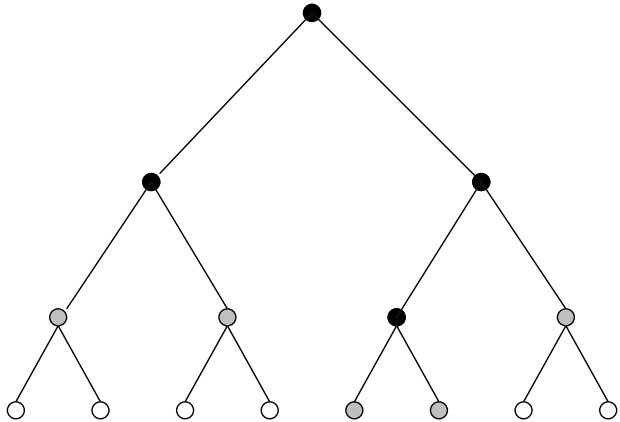}
\caption{\label{fig:eden} An Eden cluster of size $4$ on a Cayley
tree. The black sites form the cluster and the shaded sites form the
perimeter. At the next step, growth can occur at any of the $5$ shaded
perimeter sites with equal probability $1/5$. }
\end{figure}
Fig. (2) shows all possible Eden clusters of size $3$ and their
associated statistical weights.
\begin{figure}
\includegraphics[width=4in]{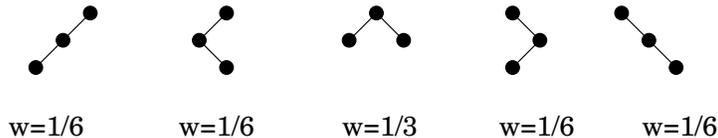}
\caption{\label{fig:eden3}All possible Eden clusters of size $3$ on a
tree and their associated statistical weights w.}
\end{figure}

\begin{figure}
\includegraphics[width=4in]{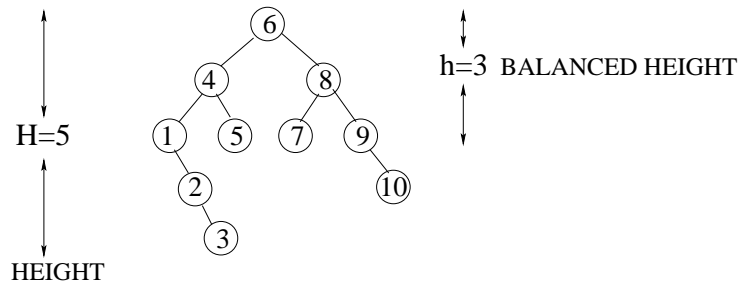}
\caption{\label{fig:bst} The binary search tree associated with the
data string $\{6,4,5,8,9,1,2,10,3,7\}$. }
\end{figure}

On the other hand, a binary search tree in computer science is
constructed by the following simple
algorithm~\cite{Knuth,Mahmoud,Review}. Imagine that we have a data
string consisting of $N$ items which are labeled by the $N$ integers:
$\{1,2,\dots,N\}$. These could be the months of the year or the names
of people etc. Let us assume that this data appears in a particular
order, say $\{6,4,5,8,9,1,2,10,3,7\}$ for $N=10$ integers. This data
is first stored on a binary tree following the simple dynamical rule:
the first item $6$ is stored at the root of the tree (see
Fig. (\ref{fig:bst})). The next item in the string is $4$. We compare
it with $6$ at the root and since $4<6$, we store $4$ in the left
daughter node of the root.  Had it been bigger than the root item $6$,
we would have stored it in the right daughter node.  The next item in
the string is $5$. We again start from the root, see that $5<6$, so we
go to the left branch. There we encounter $4$ and we find $5>4$, so we
go the right daughter node of $4$.  This process is continued till all
the $N=10$ items are assigned their nodes and we get a unique binary
search tree (BST) (see Fig. (\ref{fig:bst})) for this particular data
string $\{6,4,5,8,9,1,2,10,3,7\}$.
Usually the data arrives at a computer in random order. To study this
situation, one considers the simplest model called the `random binary
search tree' (RBST) model where one assumes that the incoming data
string can arrive in any of the $N!$ possible orders or sequences,
each with equal probability~\cite{Review}. For each of these
sequences, one has a binary tree. For example, in Fig. (4), we show the
binary trees for $N=3$ along with their associated probabilities.
\begin{figure}
\includegraphics[width=4in]{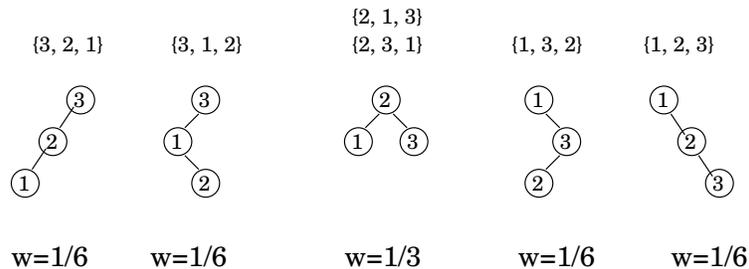}
\caption{\label{fig:bst1} All possible random binary search trees for
a data of size $N=3$ and their associated statistical weights.}
\end{figure}

Comparing Fig. (2) and Fig. (4), one sees immediately that the Eden
trees after 3 steps have exactly the same configurations and
statistical weights as the random binary search trees with data size
$N=3$.  This analogy can be easily extended to all $N$. 
The key point is that after $(n-1)$ steps there are $(n-1)$ occupied sites 
in the Eden cluster and $n$ perimeter sites (this is easy to understand as
the addition of a new occupied site eliminates one old perimeter
site while creating two new perimeter sites). 
The probability of subsequent growth
at step $n$ at any of these perimeter sites is $p_n=1/n$. Thus the
statistical weight of a cluster of $N$ sites formed by a specific
history of growth is simply $w=p_1\,p_2\,\dots p_N=1/N!$, which is the
same as in the RBST model.  Thus the Eden model on the Cayley tree is
exactly equivalent to the RBST.

Another popular search tree model is known as the `digital search
tree' (DST) which is constructed by the following
rule~\cite{Knuth,Mahmoud,Sedgewick,FS,Pittel,FR,KS,SM1}. Consider
again a binary Cayley tree each node of which can contain at most one
entry.  One starts with an empty tree and the data is stored
sequentially.  The first data item is stored in the root of the
tree. The next one arrives at the root and finding it occupied, moves
to any of the two empty daughter nodes chosen at random and occupies
that node. Then the next item arrives and again it starts at the node,
chooses any of its two daughters randomly and moves there. If the
chosen daughter is empty it occupies it. If the chosen daughter is
already occupied, it again chooses one of its two descendants at
random and moves there.  Thus at any stage, a new entry starts at the
root and performs a random walk (to the left or to the right daughter
with equal probability) down the tree till it finds an empty node and
occupies the node. Thus one obtains again a growing tree where at any
stage growth can occur at any of the perimeter sites, but now the
growth probability at a perimeter site $a$ is $p_a \propto 2^{-l_a}$
where $l_a$ is the distance of the perimeter site from the root.  The
DST is an important tree structure in computer science and has been
studied extensively. In particular, it turns out the DST is a natural
tree representation~\cite{AS,KS} of the data compression algorithm due
to Ziv and Lempel~\cite{LZ}.  Recently it was shown that a diffusion
limited aggregation model introduced by Bradley and Strenski~\cite{BS}
in physics is exactly equivalent the the DST model in computer science
and a variety of exact results were obtained by exploiting this
connection~\cite{SM1}.

The examples above illustrate a profound link between growth models
and the dynamics of search tree formation in computer science.  Note
that the two search tree models discussed above, the RBST and the DST,
can be considered as special cases of a general growth model where
growth occurs ( i.e. a new particle gets absorbed) at any of the
available perimeter sites $a$ with a growth probability $p_a\propto
c^{-l_a}$ where $c$ is a constant positive parameter and $l_a$ is the
distance of the perimeter site $a$ from the root of the tree. The RBST
(equivalently the Eden model) corresponds to $c=1$ so that all
perimeter sites have equal probability to absorb a particle. The DST,
on the other hand, corresponds to $c=2$ as discussed above. It is then
useful and interesting to study this general growth model parametrized
by $c$ and ask if there are any qualitative changes in the statistical
properties of the growth clusters as one varies the parameter $c$
continuously. Indeed, Aldous and Shields studied a continuous-time
version of this generalized growth model~\cite{AS}. Note that in the
two models discussed above time is {\em discrete} and is equal to the
number of particles in the tree. In the version of the model studied
by Aldous and Shields, time is considered {\em continuous} and growth
occurs at any of the available perimeter sites say the site $a$ with a
{\em rate} proportional to $c^{-l_a}$ where $c$ is a positive
parameter.  In this continuous-time model, the total number of
particles in the tree at time $t$ is thus a random variable, unlike in
the discrete time version. Thus, while the discrete-time model has a
constant particle number ensemble, the continuous-time model has a
constant time ensemble, much like the canonical and the grand
canonical ensemble in statistical physics.  Asymptotically at long
times, both the discrete-time and the continuous-time versions of the
model are expected behave in a similar fashion. Henceforth, we will
consider in this paper only the continuous-time version \`a la
Aldous-Shields, since it is, from a technical point of view, easier to
study than its discrete-time counterpart.

The question naturally arises whether the statistical properties of
the growing clusters in this model undergo any qualitative change of
behavior as one tunes the parameter $c$ continuously.  Indeed, Aldous
and Shields established rigorous probabilistic bounds to show that the
nature of the fluctuations (variance) in the number of particles in
the tree at time $t$ is qualitatively different for $c<\sqrt{2}$ and
$c>\sqrt{2}$. While for $c>\sqrt{2}$ the central limit theorem holds
and the total number of particles has a limiting Gaussian
distribution~\cite{AS}, for $c<\sqrt{2}$ the central limit theorem
breaks down due to the appearance of anomalously large
fluctuations. Thus, there is a sharp {\em phase transition} in the
nature of the fluctuations at a critical value $c=\sqrt{2}$. However,
the mechanism responsible for this phase transition and even the
explicit quantitative behavior of the fluctuations above, below, or at
the critical point were not easy to obtain within the rigorous
probabilistic analysis of Aldous and Shields. The principal purpose of
this paper is to provide a detailed quantitative
understanding of this rather `peculiar' phase transition.  Our method,
completely different from the original approach of Aldous and Shields,
employs a backward Fokker-Planck formalism.  The advantage of this
method is that one can obtain exact asymptotic results
explicitly. Moreover, our analysis also shows that the mathematical
mechanism behind this phase transition is similar to the phase
transitions found recently in the variance of the number of nodes
needed to store data on a $m$-ary search tree (where $m$ is the number
of branches) at the critical value
$m=26$~\cite{CH,DM1,Review,CP,FN,FFN,TH} and also in the variance of
the number of splitting events in a $D$-dimensional fragmentation
model at the critical value $D_c= \pi/{
{\sin}^{-1}\left(1/\sqrt{8}\right)}=8.69363\ldots $~\cite{DM1,Review}.

The layout of the paper is as follows. In the next Section (II), we
define the model precisely and summarize the main results. We study
here a generalized Aldous-Shields model where the growth takes place
on a Cayley tree with $m$ branches. In Section III, we derive the
evolution equations for the mean and variance of the number of
occupied sites as a function of time via a backward Fokker Planck
technique. A simple scaling analysis is then carried out to determine
the temporal growth exponents. In Section IV a more thorough analysis
of the evolution equations is provided that enables us to obtain
explicitly not just the growth exponents, but also exact expressions
for various amplitudes and prefactors that include interesting
log-periodic oscillations.  We conclude with a summary and a
discussion of open questions in the last section.

\section{The Model and the Results}

\begin{figure}
\includegraphics[width=4in]{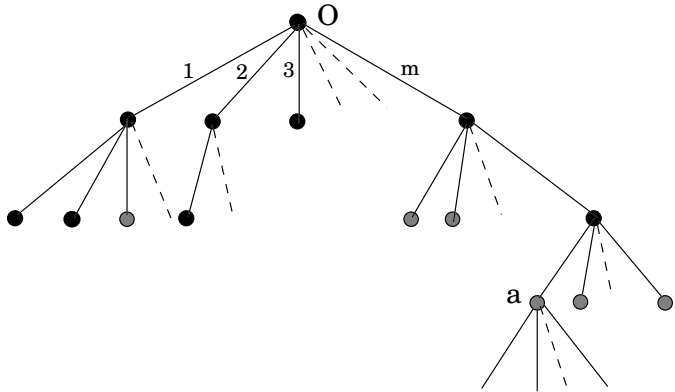}
\caption{\label{fig:mary} The growth of the Aldous-Shields model on a
tree with $m$ branches and rooted at $O$. The filled circles are
occupied nodes and the shaded ones are the perimeter nodes where
growth can occur subsequently. For example the site marked $a$ is a
typical perimeter site at a distance $l_a=3$ from the root $O$ of the
tree.}
\end{figure}

We consider a generalized Aldous-Shields model where growth occurs on
a Cayley tree (rooted at $O$) with $m$ branches (see
Fig.~\ref{fig:mary}). Aldous and Shields studied~\cite{AS} only the
binary case $m=2$. Initially the tree is empty and growth occurs in
continuous time starting from the root $O$. At any instant $t$, one
first identifies the available perimeter nodes. A node $a$ at time $t$
is a perimeter node if it is empty at $t$ but its parent node is
occupied at $t$ (see Fig.~\ref{fig:mary}). Subsequently, in a small
time interval $\Delta t$, a perimeter node $a$ either absorbs a
particle with probability $c^{-l_a}\, \Delta t$ or remains unoccupied
with probability $(1- c^{-l_a}\, \Delta t)$, where $l_a$ is the depth
of the perimeter node $a$, i.e. its distance from the root $O$. This
growth process occurs simultaneously at all the perimeter nodes. Thus,
the total number of particles $n(t)$ in the tree rooted at $O$ is
clearly not a fixed number at a given time $t$, instead it is a random
variable in the sense that the value of $n(t)$ differs from one
history of evolution to another.  We are interested in computing the
statistics of $n(t)$ at large times $t$.

In this model, we have two parameters $m$ and $c$.  It is useful to
first summarize our main results.  Using a backward Fokker-Planck
approach we derive an exact evolution equation for the generating
function,
\begin{equation}
G(\mu, t) =\langle \exp\left(-\mu\, n(t)\right)\rangle =
\sum_{n=0}^{\infty} e^{-\mu\, n} P(n,t)
\label{defG}
\end{equation}
where the angle brackets denote an average over all histories of the
evolution process and $P(n,t)$ is the probability distribution of $n$
at time $t$.  We show that $G(\mu,t)$ evolves via the equation
\begin{equation}
\frac{ dG(\mu,t)}{dt} = - G(\mu,t) + e^{-\mu}\, G^m (\mu, t/c),
\label{eqG}
\end{equation}
starting from the initial condition $G(\mu,0)=1$.  By differentiating
$G(\mu,t)$ with respect to $\mu$ and putting $\mu=0$, one can also
derive the evolution equations for all the moments of $n(t)$. The
equation (\ref{eqG}) is nonlinear and nonlocal in time for generic
values of $c$ and $m$, and is thus difficult to solve exactly, except
for the $c=1$ case when it becomes local.  However, we were able to
compute exactly the asymptotic large time behaviors of the mean and
the variance of $n(t)$ for arbitrary $m$ and $c$.  Below we present
our results for the three different cases $c=1$, $c<1$ and $c>1$
separately.
\vspace{0.3cm}

\noindent {\bf The case $c=1$:} In this case our model is precisely
the continuous-time version of the Eden model. This case $c=1$ is
exactly solvable since the evolution equation (\ref{eqG}) becomes
local in time. We solved for $G(\mu,t)$ and obtained the following
explicit result for the distribution $P(n,t)$ for all $m$ and $t$
\begin{equation}
P(n,t)=
\frac{\Gamma\left(n+\frac{1}{m-1}\right)}{\Gamma\left(\frac{1}{m-1}\right)\,
\Gamma(n+1)}\, e^{-t}\, \left[1-e^{-(m-1)\,t}\right]^{n}
\label{csol1}
\end{equation}
where $\Gamma(x)$ is the standard Gamma function. The mean number of
particles $M(t) = \langle n(t)\rangle$ increases exponentially in time
for all $m>1$,
\begin{equation}
M(t) = \frac{1}{m-1}\left[\exp\left( (m-1)t\right) -1\right].
\label{eqc1}
\end{equation} 
For the special case $m=1$ (a line with a constant rate of
deposition), $M(t)=t$ and the distribution $P(n,t)= e^{-t}
t^{n}/{n}!$, obtained from Eq. (\ref{csol1}) by taking the limit $m\to
1$, is purely Poissonian as expected.

\vspace{0.4cm}

\noindent {\bf The case $c<1$:} Since the growth rate at a perimeter
node $a$ is proportional to $c^{-l_a}$ where $l_a$ is the distance of
the node from the root $O$, it is clear that for $c<1$, farther a
perimeter node is from the root, the larger is its probability to get
occupied.  Thus the cluster grows in a rather ramified manner where
long branches grow faster than the short branches. In this case we
expect that the mean number of sites grows at least exponentially. But
since physically this case is of little interest, we do not discuss it
further in this paper.

\vspace{0.4cm}

\noindent {\bf The case $c>1$:} We now come to the physically most
relevant case $c>1$.  We show that in this case there is a sharp phase
transition in the asymptotic statistics of $n(t)$ across the
critical line $c=\sqrt{m}$ in the $(m,c)$ plane with $m>1$ and $c>1$.
We calculated exactly the asymptotic time dependence of the mean
$M(t)=\langle n(t)\rangle$ and the variance $V(t)= \langle
n^2(t)\rangle - M^2(t)$, for all values of $m>1$ and $c>1$.  We show
that while the fluctuations are normal (i.e. the variance of $n(t)$ is
proportional to its mean) for $c>\sqrt{m}$, they are anomalously large
for $c<\sqrt{m}$. Even though we have two parameters $c$ and $m$, it
turns out that the asymptotic behaviors can be described in terms of
the single growth parameter
\begin{equation}
\alpha = \frac{\ln(m)}{\ln(c)}
\label{eqalpha}
\end{equation}
where $\alpha>0$ since $c>1$. In terms of $\alpha$, the phase
transition takes phase at the critical value $\alpha_c=2$. The normal
phase for $c>\sqrt{m}$ corresponds to $\alpha<\alpha_c=2$ and the
anomalous phase for $c<\sqrt{m}$ corresponds to $\alpha>\alpha_c=2$.
\begin{figure}
\includegraphics[width=4in]{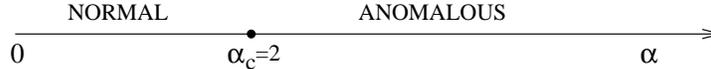}
\caption{\label{fig:aphd} The phase transition as a function of
$\alpha=\ln(m)/\ln(c)>0$.  The critical point at $\alpha_c=2$
separates the phase ($\alpha<2$) with normal fluctuations from the
phase ($\alpha>2$) with anomalously large fluctuations.}
\end{figure}

More precisely, we find that for large $t$, the mean $M(t)$ grows as a
power law (up to corrections periodic in $\ln(t)$),
\begin{equation}
M(t) \sim A\, t^{\alpha},
\label{eqm}
\end{equation}
and we provide an explicit expression for the amplitude $A$.  The
variance, on the other hand, has different behaviors for $c<\sqrt{m}$
and $c>\sqrt{m}$ or equivalently for $\alpha>2$ and $\alpha<2$.  We
show that the variance at large times $t$, again up to log-periodic
corrections, grows as
\begin{eqnarray}
V(t) &\simeq & B'\, t^{\alpha} \,\,\,\,\quad\quad \rm{for} \quad\quad
 \alpha <2 \label{eqv2}\\ &\simeq & B_c\, t^2 \ln(t) \,\,\, \rm{for}
 \quad\quad \alpha=\alpha_c=2 \label{eqvc}\\ &\simeq & B\,
 t^{2\alpha-2} \,\,\,\,\quad \rm{for} \quad\quad \alpha >2 .
 \label{eqv1}
\end{eqnarray}
We also provide exact expressions for the amplitudes $B'$, $B_c$ and
$B$.  In order to use these continuous-time results for the
discrete-time model where the `time' is same as the number of
particles, it is instructive to eliminate the explicit $t$ dependence
in the results for the variance and instead express it as a function
of the mean number of particles. Eliminating $t$ between
Eqs. (\ref{eqm}), (\ref{eqv2}), (\ref{eqvc}) and (\ref{eqv1}), we get
\begin{eqnarray}
V(t) &\sim & C'\, M(t) \,\,\,\quad\quad \rm{for} \quad\quad \alpha <2
 \label{eqvm2}\\ &\sim & C_c\, M(t) \ln(M(t))\,\,\, \rm{for}\quad
 \alpha=\alpha_c=2 \label{eqvmc}\\ &\sim & C\, M(t)^{2-2/\alpha}
 \,\,\,\,\, \rm{for} \quad\quad \alpha >2 .  \label{eqvm1}
\end{eqnarray}
Explicit expressions for the amplitudes $C'$, $C_c$ and $C$ are
likewise provided.

We thus see that for $\alpha <2$ the fluctuations of $n(t)$ about its
mean value, denoted by $\Delta n(t)=\sqrt{V(t)}$, are of order $
M^{1/2}$ as is the case for a normal Gaussian or Poisson
distribution. However for $\alpha >2$ we find that $\Delta n(t) \sim
M^{1 -{1/\alpha}}$ for large $M$. For $\alpha >2$, we have that $1-
{1/\alpha} > {1/2}$ and hence the relative fluctuations about the mean
become larger as we cross the threshold $\alpha = 2$ from below. The
phase $\alpha <2$, or equivalently $c>\sqrt{m}$ corresponds to a
region of slower growth where the central limit theorem holds and the
distribution of $n(t)$ is asymptotically normal. On the other hand,
$\alpha >2$ marks a phase where rapid growth tends to occur along a
single branch resulting in anomalously large fluctuations. Thus the
statistics of $n(t)$ in this phase is dominated by extreme
fluctuations.  The nature of this phase transition is thus very
similar to the ones recently reported in $m$-ary search
trees~\cite{CH,DM1,Review,CP,FN,FFN,TH} and a related fragmentation
model~\cite{DM1,Review}.

We end this section with a remark on the usage of the term `phase transition'.
The `phase transition' observed in this model refers to the abrupt change
of the variance (and also that of the full distribution) of the number of particles $n(t)$
in the tree as one changes the parameter $\alpha=\ln(m)/\ln(c)$ through
its critical value $\alpha_c=2$. This may not correspond to the traditional definition
of `phase transition' used in equilibrium statistical mechanics, e.g. the divergence
of a correlation length as one approaches a critical point as in second order
phase transition. The `phase transition' in the Aldous-Shields model is 
closer to the change of behavior one observes in the diffusion
of a L\'evy walker. A L\'evy walker jumps, at each step, by a random length
$l$ drawn from a power law distribution, $p(l)\sim l^{-(1+\gamma)}$ for large $l$
with $\gamma>1$ (required for normalization). It is well known\cite{BG} that 
the root mean square displacement of the particle after $n$ steps
$\sqrt{\langle R_n^2 \rangle } \sim n^{1/2}$ for large $n$ only when $\gamma>2$, i.e. one gets 
normal
diffusion and the asymptotic position of the
walker is distributed normally. On the other hand, for $1<\gamma<2$ one gets anomalously large     
diffusion, $\sqrt{\langle R_n^2\rangle} \sim n^{1/\gamma}$ for large $n$
and the asymptotic distribution of the position of the walker
is non-Gaussian. Thus 
there is a change of behavior at the critical value $\gamma_c=2$. The change of behavior
in the variance of the number of particles in the Aldous-Shields model 
at the critical value $\alpha_c=2$ is thus similar in nature to the 
the change of behavior seen for the L\'evy diffusion at $\gamma_c=2$, rather
than the standard `phase transition' observed in critical phenomena. 
  
\section{Derivation of the evolution equations}

In this section we will derive the evolution equation for the
probability distribution, and in particular the evolution equations
for the mean and the variance, of the total number of occupied sites
$n(t)$ at time $t$ in a tree rooted at the site $O$. The root $O$, by
definition, has level or depth $l_0 = 0$. The method of derivation is
based on a backward Fokker Planck formalism which involves considering
the future evolution of $n(t)$ conditioned on what happens in the
first infinitesimal time interval $(0,\Delta t)$.  Since our final aim
is to derive a recursion relation for the evolution process, it is
convenient to first derive the evolution equation for the number of
particles $n_a(t)$ in a subtree rooted, say, at any arbitrary site
$a$.  By definition, $n_a(t)$ includes the particle at the root $a$.
The number of particles in the full tree is just a special case when
the site $a$ is chosen to be the original root $O$ of the full tree.
We now count the `local' time $t$ (for this subtree) from the instant
the site $a$ becomes a potential growth site, so that, by definition,
$n_a(0)=0$.  Clearly, at any given $t$, the distribution $P(n_a,t)$ of
$n_a(t)$ depends only on $t$ and $l_a$, the depth of the site
$a$. This means that one can write
\begin{equation}
\langle F(n_a(t))\rangle = \sum_{n_a=0}^{\infty} F(n_a)\, P(n_a,t)=
f(t; l_a)
\label{inv}
\end{equation}
where $F(x)$ is any arbitrary function.

Consider now the site $a$ with its descendants $a_1,\ a_2, \cdots,
a_m$, where $a$ is a potential growth site at $t=0$. By definition $a$
is unoccupied at $t=0$ and thus $a_1,\ a_2, \cdots, a_m$ are not
potential growth sites at $t=0$.  In the first infinitesimal time
interval $(0,\Delta t)$ there are two possibilities: (i) either no
particle fills the potential growth site $a$, thus
$n_a(t)=n_a(t-\Delta t)$. This happens with probability
$1-c^{-l_a}\Delta t$. (ii) the other possibility is that the potential
growth site $a$ is filled by a particle with probability $
c^{-l_a}\Delta t$ and as a consequence the number of particles in the
subtree rooted at $a$ is increased by one and the daughter nodes
$a_1,\ a_2, \cdots, a_m$ all become potential growth
sites. Mathematically we can write the above evolution in the
following way: in the time interval $(0,\Delta t)$
\begin{equation}
n_a(t) = n_a(t-\Delta t) (1-I) + \left[1 + \sum_{i=1}^m
n_{a_i}(t-\Delta t)\right] I \label{master}
\end{equation}
where $I$ is a random variable which takes the value $1$ with
probability $c ^{-l_a}\Delta t$ and $0$ with probability $1-c
^{-l_a}\Delta t$.  Taking the expectation of Eq. (\ref{master}) with
respect to $I$ and the subsequent growth process in the remaining time
$t-\Delta t$ we obtain, upon taking the limit $\Delta t \to 0$,
\begin{equation}
\frac{d}{dt}\langle n_a(t)\rangle = c^{-l_a}\left[ 1 -\langle
n_a(t)\rangle + \sum_{i=1}^m \langle n_{a_i}(t)\rangle \right].
\end{equation}

We now use the property that the statistics of the number of particles
in a subtree rooted at level $a$ depends only on $l_a$ as encoded in
Eq. (\ref{inv}) to obtain
\begin{equation}
\frac{d}{dt}M(t;l_a) = c^{-l_a}\left[ 1 -M(t;l_a) + m M(t;l_a+1)
\right].\label{eqm1}
\end{equation}
where we have defined
\begin{equation}
M(t;l_a) = \langle n_a(t)\rangle
\end{equation}
and have used the fact that by definition $l_{a_i} = l_a +1$ if $a_i$
is a daughter of the node $a$. Note that the root $O$ of the full tree
has depth $l_O=0$. Thus the mean $M(t)=\langle n(t) \rangle$ of the
total number of particles in the full tree rooted at $O$ is simply,
$M(t)= M(t,0)$.  Thus to obtain $M(t)$, our strategy is to find the
solution $M(t,l_a)$ of Eq.  (\ref{eqm1}) for arbitrary $l_a$ and
eventually put $l_a=0$.

The next step is to notice that as one goes down a level in the tree
the growth rate is reduced by a factor of $c$ which amounts to
rescaling time by a factor of $1/c$. In the notation of
Eq. (\ref{inv}) this means that one may write
\begin{equation}
f(t; l_a + 1) = f\left(\frac{t}{c}; l_a\right).\label{scal}
\end{equation}
Next we put $l_a=0$ in Eq. (\ref{eqm1}), use the definition $M(t;l_O)
= M(t;0) = M(t)$ and also the scaling property in Eq. (\ref{scal}) to
obtain,
\begin{equation}
\frac{d}{dt}M(t) = 1 -M(t) + m M\left(\frac{t}{c}\right).\label{eqM}
\end{equation}
This equation, supplemented by the boundary condition $M(0)=0$, then
describes the evolution of the mean number of occupied sites.

An equation for the variance of $n(t)$ can be derived in a similar
fashion.  The starting point is obtained by squaring the stochastic
evolution equation (\ref{master}) and then taking the expectation over
$I$ and the evolution in the remaining time $t-\Delta t$. This yields

\begin{equation}
\langle n^2_a(t)\rangle = \langle n^2_a(t-\Delta t)\rangle (
1-c^{-l_a}\Delta t) + \left[1 + 2\langle \sum_{i=1}^m n_{a_i}(t-\Delta
t)\rangle + \langle \left(\sum_{i=1}^m n_{a_i}(t-\Delta
t)\right)^2\rangle \right]c^{-l_a}\Delta t .
\label{eqvs}
\end{equation} 
We now use the fact that the subtrees rooted at sites at the same
level are statistically independent and so
\begin{equation}
\langle n_{a_i}(t)n_{a_j}(t)\rangle = \langle n_{a_i}(t)\rangle
\langle n_{a_j}(t)\rangle \ {\rm{for}}\ i\neq j.
\end{equation}
Now defining the variance of the number of sites occupied in the tree
rooted at $0$ as
\begin{equation}
V(t) = \langle n^2(t)\rangle - \langle n(t)\rangle^2
\end{equation}
and using the scaling relation Eq. (\ref{scal}), after some elementary
 algebra, we obtain
\begin{equation}
\frac{d}{dt}V(t) = \left(\frac{d}{dt}M(t)\right)^2 -V(t) +m
V\left(\frac{t}{c}\right).
\label{eqV}
\end{equation} 
The boundary condition for this equation is clearly $V(0) = 0$.

Another way to obtain the equations for $M$ and $V$ is by deriving
directly an evolution equation for the generating function $G(\mu,t)$
of $n(t)$ defined as in Eq. (\ref{defG}).  Following exactly the same
backward Fokker-Planck strategy as used for the mean, it is
straightforward to show that $G(\mu,t)$ evolves by the nonlinear
nonlocal equation (\ref{eqG}).  The moment equations, and equations
for the higher moments, Eq. (\ref{eqM}) and Eq. (\ref{eqV}) can be
obtained by differentiating Eq. (\ref{eqG}) with respect to $\mu$ the
appropriate number of times and setting $\mu =0$ at the end.

The evolution equation (\ref{eqG}) is difficult to solve explicitly
for generic values of $c$ and $m$ since it is a nonlinear (for $m\ne
1$) and nonlocal (for $c\ne 1$) equation.  However, exact results can
be derived in a few cases that we consider below. The asymptotic
solution for the mean and variance for generic $m$ and $c$ will be
presented later in the next section.

\subsection{Exact Solution for the Eden Growth $c=1$}
The case $c=1$ corresponds to the Eden model where growth occurs at
any of the available perimeter sites with equal rates.  For $c=1$,
Eq. (\ref{eqG}) becomes local in time $t$ and can be explicitly
solved.  We find that for all $m\ge 1$
\begin{equation}
G(\mu,t) = e^{-t}\, \left[1- e^{-\mu}\,
\left(1-e^{-(m-1)t}\right)\right]^{-1/(m-1)}.
\label{c1solgf}
\end{equation}
Expanding the r.h.s. of Eq. (\ref{c1solgf}) in powers of $e^{-\mu}$ as
in Eq. (\ref{defG}), one can then read off the distribution $P(n,t)$
explicitly as in Eq. (\ref{csol1}). The mean number of particles grows
exponentially for all $m>1$ as in Eq. (\ref{eqc1}). Similarly, one can
compute the variance $V(t)$.  We find
\begin{equation}
V(t) = \frac{1}{(m-1)}\,e^{(m-1)t}\left(e^{(m-1)t}-1\right)\,.
\label{varc1}
\end{equation}
For a fixed $m$, if one takes the limit of large $t$ and large $n$
keeping the product $n\,e^{-(m-1)t}$ fixed in Eq. (\ref{csol1}), one
finds an asymptotic distribution
\begin{equation}
P(n,t) \simeq \frac{e^{-t}}{\Gamma\left(\frac{1}{(m-1)}\right)}\,
n^{-(m-2)/(m-1)}\, \exp\left[-n\, e^{-(m-1)t}\right].
\label{distc1}
\end{equation}
Thus, to the leading order, the distribution $P(n,t)$ decays
exponentially for large $n$ over a characteristic size $n^*\sim
e^{(m-1)t}$ that grows exponentially with time $t$. Interestingly, the
distribution has a sub-leading power law tail (in addition to the
leading exponential tail) $n^{-\phi}$ where the exponent
$\phi=(m-2)/(m-1)$ depends continuously on $m$.

For the special case $m=1$, where we have just a line of sites and the
particles arrive at an empty available site at a constant rate $1$, we
get from Eq. (\ref{eqc1}), $M(t)=t$. The full distribution, from
Eq. (\ref{csol1}), becomes a Poisson distribution $P(n,t)= e^{-t}
t^{n}/{n!}$ as expected.

\subsection{Exact Solution for the Digital Search Tree Growth $c=m$}

The case $c=m$ corresponds to the case where particles arrive at a
constant rate at the root $O$ and then each carries out a random walk
down the tree until it finds a free site to occupy. During its
downward journey in the tree the particle, after arriving at any
occupied site, chooses one of its $m$ descendants at random. This is
precisely the algorithm for constructing a $m$-ary digital search
tree~\cite{Knuth,Sedgewick,AS}.  If the rate at which the particles
arrive at the root $O$ is one then the total number of particles in
the tree at time $t$, $n(t)$, is clearly a random variable with Poisson
distribution
\begin{equation}
P(n=k,t) = \frac{t^k\exp(-t)}{ k!}, \label{eqPosd}
\end{equation}
where $k=0,1,2\cdots$ is a positive integer.
This yields
\begin{equation}
M(t) = t; \ V(t)= t, \label{poisson}
\end{equation}
which we see immediately are the solutions to Eq (\ref{eqM}) and Eq.
(\ref{eqV}). Furthermore we see that the generating function
$G(\mu,t)$ for a Poisson distribution is given by
\begin{equation}
G(t,\mu) = \exp\left(-t+ t \exp(-\mu)\right).
\end{equation}
It is easy to check that indeed this solves Eq. (\ref{eqG}) in the
case $m=c$.
 
\subsection{ A self-consistent scaling approach for the leading asymptotic growth of the mean and the 
variance for $c>1$ and $m>1$}

The late time asymptotic behavior of Eq. (\ref{eqM}) and
Eq. (\ref{eqV}) for $c>1$ and $m>1$ may be deduced quite simply by
making a self-consistent ansatz for the late time behavior of $M$ and
$V$. First consider Eq. (\ref{eqM}). We make the ansatz
\begin{equation}
M(t) \simeq A\, t^\alpha .
\end{equation}
Substituting this into Eq. (\ref{eqM}) we may neglect the derivative
term on the l.h.s.  and assuming that $\alpha >0$ (i.e. $c>1$),
matching the coefficients of $t^{\alpha}$ gives
\begin{equation}
\frac{m}{c^{\alpha}} - 1 = 0,
\end{equation}
which yields
\begin{equation}
\alpha = \frac{\ln(m)}{ \ln(c)}.
\end{equation}
For non-trivial tree structures we are always in the situation where
$m\geq 2$ and for the above solution to make sense we require that $c
>1$ to have a positive exponent $\alpha$. While this simple minded
scaling approach yields the correct power law growth of $M(t)\simeq
A\, t^{\alpha}$ for $c>1$, it does not provide us the value of the
amplitude $A$. To derive an exact expression for $A$, we need to solve
the full nonlocal equation (\ref{eqM}) at late times, and this will be
carried out in the next section.

Let us make a similar power law ansatz for the late time behavior of
$V(t)$
\begin{equation}
V(t) \simeq B\, t^{\beta}.
\end{equation}
Substituting this into Eq. (\ref{eqV}) and neglecting the derivative
term we obtain
\begin{equation}
- B t^{\beta} + B m \frac{t^\beta}{c^{\beta}} + A^2 \alpha^2
  t^{2\alpha -2}=0.
\label{match1}
\end{equation}
Asymptotically there are two ways to satisfy this equation.  First if
we assume a priori that $\beta > 2\alpha -2$ then the first two terms
in Eq. (\ref{match1}) must cancel leading to $c^{\beta}=m$, i.e.
$\beta = \ln(m)/\ln(c)=\alpha$. The a posteriori condition that this
solution is valid is thus $\alpha > 2\alpha -2$, which means $\alpha <
2$. The second possibility is that all three terms contribute and thus
$\beta = 2\alpha -2$. In this case we find that
\begin{equation}
B = \frac{A^2 \alpha^2}{1- \frac{m}{c^{2\alpha -2}}} = \frac{A^2
\alpha^2}{1- c^{2-\alpha}},
\label{reab}
\end{equation}
and in  obtaining the last equality in
Eq. (\ref{reab}) we have used $m=c^{\alpha}$. However for this
solution to make sense we must have that $B>0$ because $V(t)$
is clearly positive, consequently Eq. (\ref{reab}) can only hold 
when $\alpha>2$ (since $c>1$).

This simple minded scaling approach thus indicates that there is a
phase transition in the late time behavior of the variance $V(t)$ at
the critical parameter value $\alpha_c=2$. For $\alpha<2$, we have
$V(t)\sim B'\, t^{\alpha}$ where the amplitude $B'$ can not be
determined by the scaling approach. On the other hand, for $\alpha>2$
the scaling approach indicates $V(t)\sim B\, t^{2\alpha-2}$ and
moreover it provides a relationship between the amplitudes $B$ and $A$
(of the mean) via Eq. (\ref{reab}).  The critical point $\alpha_c =2$
thus separates the region of normal growth $\alpha<2$ (or equivalently
$c>\sqrt{m}$), where $V(t)\sim M(t)$, from the the region $\alpha>2$
(i.e. $c<\sqrt{m}$) where the variance grows anomalously faster $V(t)
\sim [M(t)]^{2-2/\alpha}$.  In the next section, we will see that the
analysis of the full nonlocal equations (\ref{eqM}) and (\ref{eqV})
indeed corroborates theses scaling results, and in addition produces exact
expressions for all the amplitudes.

Before proceeding to the full analysis of Eqs. (\ref{eqM}) and
(\ref{eqV}) in the next section for generic $c>1$ and $m$, it is
instructive to note that analytic progress is also possible for
Eq. (\ref{eqM}) in the case where $\alpha=\ln(m)/\ln(c)$ is a positive
integer. This includes, in particular, the critical point
$\alpha=2$. We make the following ansatz
\begin{equation}
M(t) = \sum_{k=1}^{\infty} b_n t^n,
\end{equation}
where the term $k=0$ in the above sum is omitted in order to respect
the initial condition $M(0) =0$. Matching powers of $t$ on
substituting this ansatz into Eq. (\ref{eqM}) yields
\begin{eqnarray}
b_1 &=& 1 \nonumber \\ b_{k+1} &=& b_{k} \frac{1}{k+1}
\left(\frac{m}{c^{k}}-1\right)\ {\rm for}\quad k > 1 .
\end{eqnarray}    
We thus see that if there exists a positive integer $k^*$ such that
$\alpha=\ln(m)/\ln(c)=k^*$ then $b_k = 0$ for all $k>k^*=\alpha$ and
we have found the solution to Eq. (\ref{eqM}) in these cases.  At late
times the leading order behavior is thus dominated by the term
containing $t^{\alpha}$ and we get
\begin{eqnarray}
M(t) &\sim& \frac{t^{\alpha}}{\alpha
 !}\left(\frac{m}{c^{\alpha-1}}-1\right)
 \left(\frac{m}{c^{\alpha-2}}-1\right) \cdots
 \left(\frac{m}{c}-1\right) \nonumber\\ &=& \frac{t^{\alpha}}{ \alpha
 !}\left({c}^{\alpha-1} -1\right) \left({c}^{\alpha-2} -1\right)\cdots
 \left({c}-1\right).
\label{eqmint}
\end{eqnarray}
In particular, at the critical point $\alpha=2$, we get for large $t$
\begin{equation}
M(t)\simeq \frac{(c-1)}{2}\, t^2
\label{critint1}
\end{equation}
Thus, at this special point $\alpha=2$, we have even managed to
compute the amplitude $A=(c-1)/2$ of the mean $M(t)\sim A\, t^2$
exactly.  In the case $\alpha >2$, the behavior of the variance $V(t)$
now follows immediately from Eq. (\ref{reab}).  Finally, exactly at the
critical point $\alpha =2$, we may asymptotically solve
Eq. (\ref{eqV}) with the ansatz $V = B\, t^2 \ln(t)$ to yield
\begin{equation}
B = \frac{4}{ \ln(c)}A^2 = \frac{(c-1)^2}{ \ln(c)},
\label{critint2}
\end{equation} 
where in the last line of Eq. (\ref{critint2}) we have used
$A=(c-1)/2$.

\section{General solution of the evolution equations of the Mean and the Variance}

The full solutions to the nonlocal and nonlinear differential
equations of the type in Eqs. (\ref{eqM},\ref{eqV}) are rather
difficult to obtain completely.  Here we obtain the exact asymptotic
solutions following an approach similar to the one used by Flajolet
and Richmond ~\cite{FR} in solving a class of difference-differential
equations arising in the context of digital search trees.

\subsection{ Solution for the mean $M(t)$}
We start by the analysis of Eq. (\ref{eqM}) assuming $c>1$ and
$m>1$. Taking the Laplace transform of Eq. (\ref{eqM}) we obtain
\begin{equation}
s{\tilde M}(s) = \frac{1}{ s} -{\tilde M}(s) + mc{\tilde M}(cs),
\end{equation}
where
\begin{equation}
{\tilde M}(s) = \int_0^\infty dt \ \exp(-st) M(t)
\end{equation}
and we have used the initial condition $M(0) = 0$. The above may be
written as
\begin{equation}
{\tilde M}(s) = \frac{1}{s (s+1)} +\frac{mc}{ (s+1)} {\tilde M}(cs).
\label{eqms1}
\end{equation}
Now as ${\tilde M}(s)$ should go to zero as $s\to \infty$ and we are
considering the case $c>1$, we solve Eq. (\ref{eqms1}) by iteration
finding
\begin{equation}
{\tilde M}(s) = \frac{1}{ s}\sum_{j=0}^{\infty} \frac{m^j}{
(1+s)(1+cs) \cdots (1+ c^j s)}.
\label{eqms2}
\end{equation}
Note that taking the limit $s\to 0$ is not straightforward in
Eq. (\ref{eqms2}).  This is because if we set $s=0$ in the sum on the
r.h.s of Eq. (\ref{eqms2}), the sum diverges since $m>1$.
Following~\cite{FR} we introduce the function
\begin{equation}
Q(u) = \prod_{l=0}^{\infty} \left(1+ \frac{u}{c^{l}}\right).
\label{Qdef}
\end{equation}
Thus $Q(s/c)= (1+s/c)(1+s/c^2)(1+s/c^3)\ldots$. On the other hand,
\begin{eqnarray}
Q(c^js)= \prod_{j=0}^{\infty} (1+ c^{j-l}s) &=& (1+c^j
s)(1+c^{j-1}s)\cdots (1+cs)(1+s)(1+s/c)(1+s/c^2)\ldots \nonumber \\
&=& (1+s)(1+cs)(1+c^2s)\cdots (1+c^j s) Q(s/c).
\label{Qs1}
\end{eqnarray}
Thus, one can rewrite the product $(1+s)(1+cs)\cdots(1+c^js)= Q(c^j
s)/Q(s/c)$.  Using this in Eq. (\ref{eqms2}) we get
\begin{equation}
 {\tilde M}(s) = \frac{Q({s/c})}{s}\ H(s),
\label{Ms1}
\end{equation}
where
\begin{equation}
H(s) = \sum_{j =0}^{\infty}\frac{m^j}{ Q(c^js)}.
\label{hs1}
\end{equation}
The next step is to take the Mellin transform of $H(s)$ defined as
\begin{equation}
H^*(x) = \int_0^\infty ds\ s^{x-1}\ H(s).
\label{hs2}
\end{equation}
Substituting $H(s)$ from Eq. (\ref{hs1}) in the definition in
Eq. (\ref{hs2}) we get
\begin{eqnarray}
H^*(x)& =& \sum_{j=0}^{\infty} m^j \int_0^{\infty}
\frac{s^{x-1}}{Q(c^j s)} ds \nonumber \\ &=& \sum_{j=0}^{\infty}
(mc^{-x})^{j} \int_0^{\infty} \frac{\sigma^{x-1}}{Q(\sigma)}\,d\sigma
\nonumber \\ &=& \frac{h^*(x)}{ 1-mc^{-x}},
\label{mt1}
\end{eqnarray}
where
\begin{equation}
h^*(x) = \int_0^\infty d\sigma\ \frac{ \sigma^{x-1}}{
Q(\sigma)}\label{eqh}
\end{equation}
and in evaluating the sum over $j$ we have assumed ${\rm Re}(mc^{-x}
<1)$ or equivalently ${\rm Re}(x) > \ln(m)/\ln(c)=\alpha$. We also
notice that $h^*(x)$ has no poles for ${\rm Re}(x) > 0$, and that the
poles of $1/(1- mc^{-x})$ are at $x_k = \alpha + 2\pi i k/\ln(c)$
where $k=0,\pm 1,\pm 2, \ldots$ runs over all integers.  All the poles
of $H^*(x)$ are thus to the left of the line ${\rm Re}(x) = \alpha$.

The inversion formula for the Mellin transform is given by
\begin{equation}
H(s) = {1\over 2\pi i}\int_{-i\infty + d}^{i\infty + d} dx \ H^*(x)
s^{-x},
\end{equation} 
where the above limits denote an integration up the imaginary axis to
the right of all the poles of $H^*$, therefore we chose limits with
$d>\alpha$.  The contour may be closed in the left half plane (we
assume that the integrand vanishes in the region ${\rm Re}(x)\to
-\infty$) and we can thus evaluate the inverse Mellin transform in
terms of the residues of the poles to the left of ${\rm Re}(x) \leq
\alpha$, {\em i.e}
\begin{equation}
H(s) = \sum_{\rm{poles}} {\rm Res}\left[\frac{h^*(x) s^{-x}}{1-
\exp\left (\ln(m)-x\ln(c)\right)}\right]
\label{imt1}
\end{equation}
where $\rm{Res}$ denotes the residue at the pole in question.
 
The large time behavior of $M(t)$ is determined by the small $s$
behavior of ${\tilde M}(s)$.  Now at small $s$ the dominant behavior
clearly comes from the poles $x_k= \alpha + 2\pi i k/\ln(c)$ running
up the imaginary axis, any pole coming to the left of this line of
poles will be higher order in $s$.  We evaluate the residues in
Eq. (\ref{imt1}), substitute the resulting $H(s)$ in Eq.  (\ref{Ms1})
and then take the limit $s\to 0$ to obtain the following asymptotic
result
\begin{equation}
{\tilde M}(s) \sim \frac{1}{ s^{\alpha+1}\ln(c)}\left[ h^*(\alpha) +
\sum_{k\neq 0} h^*(\alpha + 2\pi i k/\ln(c))s^{\frac{2\pi i k}{
\ln(c)}}\right],\label{eqHs}
\end{equation}   
where we have used the fact $Q(0)=1$.  Note that from Eq. (\ref{eqh})
and Eq. (\ref{Qdef}), we have
\begin{equation}
h^*(\alpha)= \int_0^{\infty}
\frac{\sigma^{\alpha-1}\,d\sigma}{(1+\sigma)(1+\sigma/c)(1+\sigma/c^2)\ldots}
= \frac{\pi}{ \sin(\pi \alpha)} \prod_{k=1}^{\infty} \frac{1-
c^{\alpha-k}}{1-c^{-k}},
\label{rameq}
\end{equation} 
where the last equality follows from an identity due to
Ramanujan~\cite{ram}.  Note that this identity explicitly shows that
the function $h^*(x)$ has simple poles at the negative integers and
zero but no poles for ${\rm Re}(x) >0$ as was stated before.

To extract the leading asymptotic behavior of $M(t)$ for large $t$,
let us first divide the Laplace transform ${\tilde M}(s)$ into two
parts, ${\tilde M}(s)= {\tilde M_p}(s) +{\tilde M_l}(s)$ where
${\tilde M_p}(s)$ denote the first term on the r.h.s. of
Eq. (\ref{eqHs}) and ${\tilde M_l}(s)$ corresponds to the remaining
sum over $k\ne 0$.  Subsequently the inverse Laplace transform $M(t)=
M_p(t) + M_l(t)$ can also be divided into two parts.  The term
${\tilde M_p}(s)$ has a pure algebraic form, thus its inverse $M_p(t)$
has a pure power law growth,
\begin{equation}
M_p(t) \sim A\ t^{\alpha},
\label{Mpt1}
\end{equation}
where the constant $A$ can be evaluated as follows. If $M_p(t)$ has
the form in Eq. (\ref{Mpt1}), its Laplace transform is ${\tilde
M_p}(s)= A\, \Gamma(1+\alpha)\, s^{-(1+\alpha)}$. Comparing this with
the first term in Eq. (\ref{eqHs}) gives
\begin{equation}
A = \frac{h^*(\alpha)}{\ln (c) \Gamma(1+\alpha)} = \frac{\pi}{ \ln(m)
\Gamma(\alpha) \sin(\pi \alpha)} \prod_{k=1}^{\infty}
\frac{1-c^{\alpha-k}}{1-c^{-k}},
\label{A1}
\end{equation}
where we have used $\Gamma(1+\alpha)= \alpha \Gamma(\alpha)$, the
definition $\alpha= \ln(m)/\ln(c)$ and the explicit form of
$h^*(\alpha)$ from Eq. (\ref{rameq}). Here we note that when $\alpha$
is an integer, it can be verified that Eq. (\ref{A1}) agrees with
Eq. (\ref{eqmint}) derived for discrete values of $\alpha$ in the
previous section.

The second contribution to ${\tilde M}(s)$, $\tilde{M}_l(s)$ is given
by a Fourier series in $\ln(s)$. The inverse Laplace transform of this
term is difficult to obtain fully but it is easy to see that it gives
rise to a late time behavior of the form
\begin{equation}
M_l(t) \sim {A\, t^{\alpha}} g\left(\ln(t)\right)
\end{equation}
where $g(x)$ is a periodic function of $x$.  The final asymptotic
result for large $t$ is thus
\begin{equation}
M(t)=M_p(t)+M_l(t) \simeq {A\, t^{\alpha}}\left[1+
g\left(\ln(t)\right)\right]. \label{eqmtb}
\end{equation}
This exact result thus not only confirms the dominant power-law
scaling predicted in section (III) up to log-periodic oscillations,
but also provides an explicit formula for the amplitude $A$ as in
Eq. (\ref{A1}). For example, let us consider the binary case
$m=2$. For the case, when $c=1$, the formula in Eq. (\ref{A1}) gives
$A=1$, thus $M(t)\simeq t$ for large $t$.  On the other hand, for
$m=2$, when $c=\sqrt{2}$ (the critical point), one can show from
Eq. (\ref{A1}) that $A=(\sqrt{2}-1)/2$ and $M(t) \simeq (\sqrt{2}-1)
t^2 /2$ for large $t$.

\subsection{ Solution for the variance $V(t)$}

We now examine the asymptotic behavior of the variance $V(t)$ for
large $t$ using a similar formalism.  The evolution equation
(\ref{eqV}) for the variance $V(t)$ is similar to that for the mean
$M(t)$ in Eq. (\ref{eqM}) except that the source term in
Eq. (\ref{eqV}) is $(dM/dt)^2$, different from the source term $1$ in
Eq. (\ref{eqM}).  Solution of Eq. (\ref{eqV}) thus requires an
explicit knowledge of how $M(t)$ behaves with time.  Taking the
Laplace transform, ${\tilde V}(s)=\int_0^{\infty} V(t) e^{-st}\, dt$
in Eq. (\ref{eqV}) and using $V(0)=0$ we obtain
\begin{equation}
s{\tilde V}(s) = S(s) -{\tilde V}(s) + mc{\tilde V}(cs),
\label{eqS}
\end{equation}
where
\begin{equation}
S(s) = \int_0^\infty dt \exp(-st) \left( \frac{d M}{dt} \right)^2.
\label{ss}
\end{equation}
Rearranging Eq. (\ref{eqS}) gives
\begin{equation}
{\tilde V}(s)= \frac{S(s)}{1+s} + \frac{mc}{1+s}\, {\tilde V}(cs)
\label{vs1}
\end{equation}
which can be iterated to yield
\begin{equation}
{\tilde V}(s) = \sum_{j=0}^{\infty} \frac{(mc)^j}{(1+s)(1+cs)\cdots
(1+c^js)}\, S(c^j s).
\label{vs2}
\end{equation}
Using $m=c^{\alpha}$ and the function $Q(u)$ defined in
Eq. (\ref{Qdef}) we can rewrite Eq. (\ref{vs2}) as
\begin{equation}
{\tilde V}(s) = Q(s/c) H_1(s)
\label{vs3}
\end{equation}
where
\begin{equation}
H_1(s) = \sum_{j=0}^{\infty} \left(c^{1+\alpha}\right)^{j}
\frac{S(c^js)}{Q(c^j s)}.
\label{h1s1}
\end{equation}

The next step is to take the Mellin transform
$H_1^*(x)=\int_0^{\infty} H_1(s) s^{x-1} ds$ of Eq. (\ref{h1s1}) which
gives, after a change of variable $c^j s\to s$ in the integration
\begin{equation}
H_1^*(x) = \sum_{j=0}^{\infty} \left(c^{1+\alpha-x}\right)^{j}
\int_0^{\infty} \frac{S(s)}{Q(s)} s^{x-1} ds .
\label{h1x1}
\end{equation}
Let us first assume that the integral
\begin{equation}
h_1^*(x) = \int_0^{\infty} \frac{S(s)}{Q(s)} s^{x-1} ds
\label{h1x2}
\end{equation}
exists (the conditions for which will be stated later). Then, for
${\rm Re}(x)>1+\alpha$, the geometric sum in Eq. (\ref{h1x1})
converges (since $c>1$) and we get
\begin{equation}
H_1^*(x) = \frac{h_1^*(x)}{1- c^{1+\alpha-x}}.
\label{h1x3}
\end{equation}
Inverting this Mellin transform we get
\begin{equation}
H_1(s) = \frac{1}{2\pi i} \int_{-i\infty +d}^{i \infty +d}
\frac{h_1^*(x)}{1- c^{1+\alpha-x}} s^{-x} dx = \sum_{\rm poles} {\rm
Res}\left[ \frac{h_1^*(x)}{1- c^{1+\alpha-x}} s^{-x}\right],
\label{h1s2}
\end{equation}
where the poles are at $x_k= 1+\alpha - 2\pi ik/{\ln(c)}$ with
$k=0,\pm1,\pm2,\ldots$. In Eq. (\ref{h1s2}) the integration is along
the imaginary axis to the right of all the poles and then we close the
contour over the left half plane. Evaluating the residues and
substituting the results in Eq. (\ref{vs3}) we get
\begin{equation}
{\tilde V}(s) = \frac{Q(s/c)}{s^{1+\alpha}
\ln(c)}\left[h_1^*(1+\alpha) + \sum_{k\ne 0} h_1^*\left(1+\alpha-2\pi
ik/\ln(c)\right)\right],
\label{vs4}
\end{equation} 
where $h_1^*(x)$ is given by Eq. (\ref{h1x2}) assuming that it exists.

We now need to invert the Laplace transform in Eq. (\ref{vs4}) to
evaluate $V(t)$. For large $t$, as usual, the dominant contribution
will come from the small $s$ behavior of ${\tilde V}(s)$.  Using
$Q(0)=1$ and assuming $h_1^*(1+\alpha-2\pi ik/\ln(c))$ exists for all
$k=0,\pm 1, \pm 2\ldots$, it is clear from the small $s$ behavior of
${\tilde V}(s)$ in Eq. (\ref{vs4}) that for large $t$
\begin{equation}
V(t) \simeq B'\, t^{\alpha}[1+ G(\ln(t))],
\label{vta1}
\end{equation}
where $G(x)$ is a periodic function in $x$ and the amplitude $B'$ can
be read off as
\begin{equation}
B'= \frac{h_1^*(1+\alpha)}{\Gamma(1+\alpha)\ln(c)};\,\,\, {\rm
where}\,\,\,\, h_1^*(1+\alpha)=\int_0^{\infty} \frac{S(s)}{Q(s)}
s^{\alpha} ds.
\label{b'1}
\end{equation}    
  
Having obtained the results in Eqs. (\ref{vta1}) and (\ref{b'1}), we
need to investigate when they are valid. These results are valid as
long as the integral $h_1^*(1+\alpha)$ in Eq. (\ref{b'1}) exists. The
existence of this integral depends on the small $s$ behavior of the
source function $S(s)$ defined in Eq. (\ref{ss}). Using the asymptotic
behavior of $M(t)$ from Eq. (\ref{eqmtb}) we find that for large $t$
\begin{equation}
\left(\frac{dM}{dt}\right)^2 \simeq A^2\alpha^2 t^{2\alpha-2} \left[1+
g_1(\ln(t)\right],
\label{dmdt2}
\end{equation}
where $g_1(x)$ is a periodic function in $x$. Substituting this large
$t$ behavior of $(dM/dt)^2$ in Eq. (\ref{ss}), it follows that, in the
case $\alpha < 1/2$, the integral converges to a nonzero constant as
$s\to 0$.  On the other hand, for $\alpha>1/2$, the integral diverges
as $S(s)\simeq A^2\alpha^2 \Gamma(2\alpha-1) s^{-(2\alpha-1)}$ as
$s\to 0$. Up to the log-periodic oscillations, the leading behavior of
$S(s)$ for small $s$ can be summarized as follows
\begin{eqnarray}
S(s) &\simeq & C_0 s^{-(2\alpha-1)}\,\,\,\quad {\rm for}\quad\quad
\alpha>1/2
\label{aless} \\
&\simeq & -\ln(s) \,\,\,\quad {\rm for}\quad\quad \alpha=1/2
\label{aequal} \\ &\simeq & A_1 \,\,\,\quad\quad {\rm for}\quad\quad
\alpha<1/2
\label{amore}
\end{eqnarray}
where
\begin{equation}
C_0= A^2\alpha^2 \Gamma(2\alpha-1)
\label{Ceq}
\end{equation}
is a positive constant for $\alpha>1/2$. Also, $A_1= \int_0^{\infty}
(dM/dt)^2 dt$ is a constant that depends on the full form of $M(t)$
and not just on its asymptotic behavior since for $\alpha<1/2$ the
integral is convergent.  Substituting this small $s$ behavior of
$S(s)$ into the integral giving $h_1^*(1+\alpha)$ in Eq. (\ref{b'1})
and using $Q(0)=1$, it is clear that the integral exists (no
divergence from the small $s$ limit) only for $\alpha<2$. For
$\alpha>2$, the integral does not exist since the integrand for small
$s$ scales as $s^{1-\alpha}$. Thus the results in Eqs. (\ref{vta1})
and (\ref{b'1}) hold only for $\alpha<2$.

For $\alpha>2$, the above analysis breaks down and we need to employ a
different method. We now go back to our starting equations (\ref{vs3})
and (\ref{h1s1}). It turns out that for $\alpha>2$, we can actually
extract the leading small $s$ behavior directly from these two
equations. We directly substitute in Eq. (\ref{h1s1}) the leading
small $s$ behavior of $S(s)\simeq C_0 s^{-(2\alpha-1)}$ from
Eq. (\ref{aless}) where $C_0=A^2\alpha^2
\Gamma(2\alpha-1)$. Additionally we use $Q(0)=1$. Eqs. (\ref{vs3}) and
(\ref{h1s1}) then yield in the $s\to 0$ limit
\begin{equation}
{\tilde V}(s) \simeq \frac{C_0}{s^{2\alpha-1}} \sum_{j=0}^{\infty}
\left(c^{2-\alpha}\right)^j = \frac{C_0}{s^{2\alpha-1} (1-
c^{2-\alpha})}
\label{vs5}
\end{equation}
where we have used $\alpha>2$ which ensures that the sum in
Eq. (\ref{vs5}) is convergent.  Inverting the Laplace transform, we
then get the large $t$ behavior of $V(t)$ for $\alpha>2$
\begin{equation}
V(t) \simeq B\, t^{2\alpha-2};\,\,\,\, {\rm where}\,\,\,\, B=
\frac{\alpha^2 A^2}{1-c^{2-\alpha}},
\label{vs6}
\end{equation}
where the constant $A$ is given in Eq. (\ref{A1}). Note that this
result in Eq. (\ref{vs6}) for $\alpha>2$ is in perfect agreement with
the self-consistent scaling approach used in Section-III.

At the critical point $\alpha=2$, the analysis is more
delicate. However, from the scaling approach of Section-III, we
already know that for large $t$, $V(t)\simeq B_c\, t^2 \ln(t)$ with
$B_c= (c-1)^2/\ln (c)$ as in Eq. (\ref{critint2}). Thus, the
asymptotic behavior of the variance $V(t)$ can be summarized as
\begin{eqnarray}
V(t) &\simeq & B'\, t^{\alpha} \,\,\,\quad \rm{for} \quad \alpha <2 \\
 &\simeq & B_c\, t^2 \ln(t) \,\,\,\quad \rm{for} \quad
 \alpha=\alpha_c=2 \\ &\simeq & B\, t^{2\alpha-2} \,\,\,\,\,\quad
 \rm{for} \quad \alpha >2 , \label{varat}
\end{eqnarray}
where the three amplitudes are given by
\begin{eqnarray}
B'&=& \frac{1}{\Gamma(1+\alpha)\ln(c)}\int_0^{\infty}\frac{S(s)}{Q(s)}
s^{\alpha} ds \nonumber \\ B_c &=& (c-1)^2/{\ln (c)} \nonumber \\ B
&=& \frac{\alpha^2 A^2}{1-c^{2-\alpha}} \label{bs}
\end{eqnarray}
where $A$ is given in Eq. (\ref{A1}). Note that computing the
amplitude $B'$ explicitly requires an integration over the full source
function $S(s)$ which is not so easy.  Eliminating the time $t$
between $M(t) \simeq A\, t^{\alpha}$ and $V(t)$ in Eq. (\ref{varat}),
one can express the variance $V$ as a function of the mean $M$ for
large $M$ as in Eqs. (\ref{eqvm2}), (\ref{eqvmc}) and (\ref{eqvm1})
and one can read off the constants $C'$, $C_c$ and $C$ in terms of
$B'$, $B_c$ and $B$ and the amplitude $A$ of the mean given in
Eq. (\ref{A1}).

Let us end this section with a remark on the mathematical mechanism
responsible for the phase transition in the variance of the number of
particles in this Aldous-Shields model. We note that the exact
evolution equations (\ref{eqM}) and (\ref{eqV}) respectively for the
mean and the variance are very similar---they are both linear and
nonlocal in time, the only difference is in the source term. For the
mean $M(t)$ in Eq. (\ref{eqM}), the source term is a constant $1$ (the
first term on the r.h.s of Eq. (\ref{eqM})). On the other hand, for
the variance, the source term $(dM/dt)^2$ in Eq. (\ref{eqV}) depends
on the evolution of the mean. Thus, the mean feeds into the variance
equation as an external source term leading to a competition between
the growth induced by this external source term and the growth induced
internally by the remaining two terms on the r.h.s of
Eq. (\ref{eqV}). This competition between the external and the
internal source is finally responsible for the phase transition in the
asymptotic growth of $V(t)$. For $\alpha<2$, the internal source term
wins out and the variance grows similarly as the mean, $V(t) \sim M(t)
\sim t^{\alpha}$, leading to the normal phase. On the other hand, for
$\alpha>2$, the external source term wins out leading to a faster
growth $V(t)\sim t^{2\alpha-2}\sim [M(t)]^{2-2/\alpha}$ characterizing
anomalously large fluctuations. We note that a similar mechanism
namely a ``{\it competition between the internal source and the external
driving}" was shown to be responsible for phase transitions in
fluctuations in a class of fragmentation problems studied
recently\cite{DM1,Review}. In these fragmentation problems, it was
shown that the mean and the variance evolved via similar looking
equations\cite{DM1,Review}, except that they differed in their
respective source terms---the variance equation had a source term
driven by the mean, in much the same way as in the Aldous-Shields
model discussed here. Thus, it seems that this phase transition in
fluctuations is quite generic as it occurs in a large class of
problems and the mathematical mechanism responsible for it is as
identified above.

\section{Summary and Conclusion}

In this paper we have studied analytically a growing tree model
introduced by Aldous and Shields.  In this model, growth occurs in
continuous time. One starts at $t=0$ with an empty Cayley tree with
$m$ branches rooted at $0$ and the tree grows, starting from the root
site, by absorbing particles in continuous time. Each site can occupy
at most one particle. At a given instant $t$, growth can occur only at
the perimeter sites with a rate $c^{-l}$ where $c$ is positive
parameter and $l$ is the distance of the perimeter site from the root
of the tree.  For $c=1$ this model is isomorphic to a continuous-time
Eden model on a tree and also corresponds to the random binary search
tree problem in computer science.  For $c=2$ this model corresponds to
the digital search tree problem in computer science.

We have introduced a backward Fokker-Planck approach that enabled us
to study analytically the statistics of the total number of particles
$n(t)$ in the tree at large time $t$. We have shown that at large $t$,
while the mean number of particles grows as a power law in time,
$M(t)\simeq A\, t^{\alpha}$ with $\alpha=\ln(m)/\ln(c)$ for all $c>1$,
the variance $V(t)$ of the number of particles has two different
behaviors depending on the value of the parameter $\alpha$. While for
$\alpha<2$ $V(t)\sim M(t)$ for large $t$, for $\alpha>2$ the variance
grows anomalously quickly: $V(t)\sim [M(t)] ^{2-2/\alpha}$.  We have
identified the mathematical mechanism behind this phase transition at
the critical value $\alpha_c=2$ and shown that it is qualitatively
similar to the phase transitions recently encountered in a search tree
problem and also in a related fragmentation problem. Essentially, for
$\alpha<2$, the typical value of $n(t)$ grows in the same way as the
average and the distribution is asymptotically normal whereas for
$\alpha>2$, the typical value does not grow the same way as the average and
the distribution is characterized by large fluctuations caused
by the faster growth of a single branch of the tree.

We obtained detailed analytical results for the first two moments of
the number of particles for generic values of the parameter
$\alpha$. However, we were able to calculate the full asymptotic
distribution of the number of particles only for two specific values
of $\alpha$, namely for $\alpha=1$ ($c=m$) and $\alpha\to \infty$
($c=1$). Fortunately these two representative values, where an exact
solution is possible, fall respectively on either side of the critical
point $\alpha_c=2$. Our exact solution shows that for $\alpha=1$
($<\alpha_c=2$) the distribution $P(n,t)$ is Poisson and hence is
asymptotically normal for large $n$. On the other hand for $\alpha\to
\infty$, the asymptotic distribution is certainly non-Gaussian,
$P(n,t) \sim n^{-\phi} \exp[-n e^{-(m-1)t}]$ where the exponent
$\phi=(m-2)/(m-1)$ depends on $m$. The calculation of the distribution
for other values of $\alpha$ remains a challenging problem.

While we have studied this growth model on a tree because of its
connections to the search tree problems as mentioned in the
introduction, it is of general interest to study this growth problem
on a regular Euclidean lattice, e.g. on a hyper-cubic lattice in $d$
dimensions. In this lattice model, the cluster will grow similarly
from a seed site at the origin. At a given instant, growth can occur
at any of the available surface sites with a rate $c^{-r}$ where $r$
is the Euclidean distance of the surface site from the origin. One can
then investigate the statistics of the total number of particles in
the cluster after time $t$. It is easy to make a scaling argument for
the growth of the mean number of particles $M(t)$. Assuming that the
cluster is compact with a typical radius $R(t)$ at time $t$, we have
$M(t)\sim [R(t)]^{d}$. Also, the mean number of surface sites
$N_p(t)\sim [R(t)]^{d-1}$. By the growth rule, $dM/dt \sim N_p(t)
c^{-R(t)}$ for large $t$. This predicts $R(t)\sim \ln(t)$ for large
$t$ and hence the mean number of particles grows very slowly as
$M(t)\sim [\ln (t)]^{d}$ for large $t$. An interesting open question
for future studies is whether, in finite dimensional lattice models, 
the variance exhibits a  phase transition, similar to that seen  on the tree, 
for some critical value of the parameter $c$?

\end{document}